\documentclass[prd,aps,nofootinbib,notitlepage,preprintnumbers
]{revtex4}
\usepackage{graphicx,epsf,amsmat
h,amsfonts,amssymb,amsbsy}
\usepackage{epsfig}
\newcommand{\mat}{\left ( \begin{array}}
\newcommand{\emat}{\end{array} \right )}
\newcommand{\vect}{\left ( \begin{array}{c}}
\newcommand{\evect}{\end{array} \right )}

\def\cP{\mathcal P}

\begin{document}


\title{
 Composite operator approach 
 to dynamical mass generation in the (2+1)-dimensional Gross-Neveu model
}
\author{T. G. Khunjua $^{1}$, K. G. Klimenko $^{2}$, and R. N. Zhokhov $^{2,3}$ }

\affiliation{$^{1}$ The University of Georgia, GE-0171 Tbilisi, Georgia}
\affiliation{$^{2}$ State Research Center
of Russian Federation -- Institute for High Energy Physics,
NRC "Kurchatov Institute", 142281 Protvino, Moscow Region, Russia}
\affiliation{$^{3}$  Pushkov Institute of Terrestrial Magnetism, Ionosphere and Radiowave Propagation (IZMIRAN),
108840 Troitsk, Moscow, Russia}

\begin{abstract}
Using a nonperturbative approach based on the Cornwall-Jackiw-Tomboulis (CJT) effective action 
$\Gamma(S)$ for composite operators, 
the phase structure of the simplest massless (2 + 1)-dimensional Gross-Neveu model is 
investigated. We have calculated $\Gamma(S)$ in the first order of the bare coupling constant $G$
and have shown that there exist three different specific dependences of  
$G\equiv G(\Lambda)$ on the cutoff parameter $\Lambda$, and in each case 
the effective action and its stationarity equations have been obtained. The solutions of these 
equations correspond to the fact that three different masses of fermions can arise dynamically 
and, respectively, three different nontrivial phases can be observed in the model. 
\end{abstract}
\maketitle
\section{Introduction}

In the last few decades, much attention has been paid to the study of (2 + 1)-dimensional (D) 
field theory models, which can be used to predict and study physical phenomena of planar 
nature such as quantum Hall effect, high-temperature superconductivity, low-energy 
graphene physics, etc. To a fairly 
large extent, these phenomena are usually considered within the framework of models with 
a four-fermion interaction \cite{Semenoff,Shovkovy,Gusynin,Chaves,Mesterhazy,Vshivtsev,Khudyakov,Kanazawa,Gomes,Ebert}. One of the 
reasons is that in these models the spontaneous symmetry breaking occurs dynamically, i.e. 
without taking into account additional scalar Higgs bosons.
Moreover, despite the perturbative nonrenormalizability of these (2 + 1)-dimensional models,
in the framework of nonperturbative approaches such as 
the large-$N$ technique, etc., they are renormalizable \cite{Rosenstein}. And just using this 
nonperturbative $1/N$ approach, spontaneous symmetry breaking and the associated dynamic 
generation of the fermion mass effects were investigated in the simplest (2 + 1)-D model 
with four-fermion interaction (called the Gross-Neveu (GN) model  \cite{GN}). \footnote{Its Lagrangian is presented 
below in Eq. (\ref{n1}). } In particular, it was shown, e.g., in Ref. \cite{Modugno} that at zero 
temperature and zero
chemical potential (as well as at fixed value of the cutoff parameter $\Lambda$) in this 
(2+1)-D GN model a phase with dynamical chiral symmetry breaking is occured only for sufficiently 
large (positive) values of the bare coupling constant $G\equiv G(\Lambda)$. For a rather weak interaction the 
symmetric phase is realized in the model, and it is not an asymptotically free one.
(In contrast, the (1+1)-D GN model \cite{GN} is an asymptotically free and dynamic generation 
of the fermionic mass occurs there for arbitrary values of bare coupling constant.)
Qualitatively the same properties of this (2+1)-D GN model one can observe in terms of variational 
optimized expansion technique \cite{Klimenko:1993} and other nonperturbative variational 
approarches \cite{Kneur:2007}, etc.

Unlike the aforementioned papers, we investigate the (2+1)-D GN model (\ref{n1}) within the framework 
of another nonperturbative approach based on the effective action for composite operators. Originally, 
the approach was proposed in the paper by Cornwall-Jackiw-Tomboulis (CJT) \cite{CJT} when 
considering mainly a scalar $\phi^4$-field model, etc. Then in a series of papers 
\cite{Peskin,Casalbuoni,Dorey,Rochev,Appelquist} the CJT effective action for
composite operators method has been extended to quantum field theory models with fermions. As a 
result, a nonperturbative method has emerged for calculating various multi-fermion 
Green's functions based on functional equations of the Dyson-Schwinger type. Moreover, 
in this CJT effective action approach it is possible to 
investigate the possibility of dynamical generation of the fermion mass and chiral 
symmetry breaking, etc, as it was demonstrated in the framework of the (1+1)-D GN model in Ref. 
\cite{Dorey}. Note that in the last case, i.e. in (1+1)-D, the results of the CJT effective 
action studies of the model are qualitatively the same as in the large-$N$ expansion technique.
We study phase structure of the (2+1)-D GN model using the CJT effective action calculated up to 
a first order in the coupling constant $G$. It turns out that in this case, in contrast to (1+1)-D 
GN model, the CJT approach predicts a much richer phase structure compared to the result obtained
with other generally accepted nonperturbative methods, i.e. large-$N$ and optimized expansion techniques, etc.

It is well-known that different states or phases of various (2+1)-dimensional condensed matter 
systems such as graphene, etc. can be described in the framework of the same relativistic 
model, when corresponding nonzero order parameters appear in it under the influence of 
external conditions (temperature, chemical potentials, etc), or when free model parameters 
such as coupling constants, etc are changed. It is important to emphasize that if the system 
(or relativistic model) consists of massless fermions, then bilinear fermion combinations 
(or simply the mass terms of the model Lagrangian), which arise spontaneously, are usually 
used as order parameters for one or another phase of the system. In (2+1)-dimensions, 
and in particular in the models used to describe the physics of graphene, there 
is a fairly rich selection of fermionic masses (see, e.g., in Refs. 
\cite{Gusynin,Mesterhazy,Mudry}). However, for simplicity, in 
the present paper, using the CJT effective action method, we investigate the possibility of 
the dynamic occurrence of only four of them within the framework of the simplest (2+1)-D GN 
model (\ref{n1}). (The dynamical emergence of other mass terms can be analized in a similar 
way.) The dynamic appearance of each of the masses corresponds to the fact 
that a spontaneous breakdown of one or another discrete symmetry (or their combination) 
occurs, and that the system has passed into a certain well-defined phase. For example,
in one of these phases, the usual Dirac mass of fermions is generated dynamically,
and spontaneous breaking of discrete chiral symmetries $\Gamma^5$ and $\Gamma^3$ 
(see definition in Eq. (\ref{n4}))  of the original Lagrangian (\ref{n1}) occurs. The other 
phase is characterized by a dynamical generation of a mass order parameter called the Haldane 
fermion mass which violates the spatial parity $\cal P$, etc. Which of these phases is 
realized in the system depends on the behavior of the bare coupling constant $G(\Lambda)$ vs 
cutoff parameter $\Lambda$. 
 
The paper is organized as follows. Section II presents the $N$-flavor massless 
(2+1)-dimensional Gross-Neveu model symmetric with respect to chiral $\Gamma^5$ and $\Gamma^3$ 
as well as parity $\cP$ discrete transformations. 
In addition, the CJT effective action $\Gamma(S)$ of the composite bilocal and bifermion operator 
$\overline\psi (x)\psi (y)$ is constructed here, which is actually the functional of the full 
fermionic propagator $S(x,y)$. In real situations, the propagator is a translation invariant 
solution of the stationary Schwinger-Dyson-type equation of the CJT effective action. 
In this section, the unrenormalized expression for $\Gamma(S)$ is obtained up to a first order 
in the coupling constant $G$. Based on this 
expression, we show in section III (section IV) that there exists a renormalized, i.e. 
without ultraviolet divergences, solution of the Schwinger-Dyson-type equation for the propagator 
corresponding to fermions with dynamically arising nonzero Dirac mass 
(fermions with Haldane mass). Finally, in Section V the possibility for dynamical generation 
of the mass term of the form $\left (m_5\overline\psi i\gamma^5\psi
+m_3\overline\psi i\gamma^3\psi\right )$ is established in the system.
In each of these phases of the model, the renormalization and dynamic appearance of the 
fermion mass occurs due to 
a specific behavior of the bare coupling constant $G$ vs cutoff parameter (each phase has its own
behavior of $G$). 

\section{(2+1)-dimensional GN model and its CJT effective action}

\subsection{Model, its symmetries, etc}

The Lagrangian of the simplest massless and $N$-flavor (2+1)-dimensional GN model under 
consideration has the following form
\begin{eqnarray}
 L=\overline \psi_k\gamma^\nu i\partial_\nu \psi_k&+& \frac {G}{2N}\left
(\overline \psi_k\psi_k\right )^2,
\label{n1}
\end{eqnarray}
where for each $k=1,...,N$ the field $\psi_k\equiv \psi_k(t,x,y)$ is a (reducible) four-component Dirac spinor
(its spinor indices are omitted in Eq. (\ref{n1})), $\gamma^\nu$ 
($\nu=0,1,2$) are 4$\times$4 matrices acting in this four-dimensional spinor space (the algebra
of these $\gamma$-matrrices and their particular representation used in the present paper
is given in Appendix \ref{ApC}, where the matrices $\gamma^3,\gamma^5$ and $\tau=-i\gamma^3\gamma^5$
are also introduced), and the summation over repeated $k$- and $\nu$-indices is assumed in Eq. 
(\ref{n1}) and below. The bare coupling constant $G$ has a dimension of [mass]$^{-1}$. 
As discussed, e.g., in Refs. \cite{Gusynin,Ebert}, at $N=2$ the model provides a fairly 
good description of the low-energy physics of graphene in the continuum limit. But  
we consider the $N$-flavor variant of the GN model in order to compare its phase structure 
obtained in the framework of the CJT effective approach with the results of the large-$N$ 
investigation \cite{Rosenstein}. So the model (\ref{n1}) is invariant under $U(N)$ (but in the 
following this symmetry remains intact). More important for us is that the 
Lagrangian is invariant under three discrete transformations, two of them are the so-called 
chiral transformations $\Gamma^5$ and $\Gamma^3$,
\begin{eqnarray}
 \Gamma^5:&~~&\psi_k(t,x,y)\to  \gamma^5\psi_k(t,x,y);~~
 \overline\psi_k(t,x,y)\to  -\overline\psi_k(t,x,y)\gamma^5,\nonumber\\
 \Gamma^3:&~~&\psi_k(t,x,y)\to  \gamma^3\psi_k(t,x,y);~~
 \overline\psi_k(t,x,y)\to  -\overline\psi_k(t,x,y)\gamma^3.
 \label{n4}
\end{eqnarray}
The rest one is the space reflection, or parity, transformation $\cP$ under which $(t,x,y)\to 
(t,-x,y)$ \footnote{In 2+1 dimensions, parity corresponds to inverting only one 
spatial axis \cite{Semenoff,Appelquist}, since the inversion of both axes is equivalent to 
rotating the entire space by $\pi$.} and
\begin{eqnarray}
 \cP:~~\psi_k(t,x,y)\to  \gamma^5\gamma^1\psi_k(t,-x,y);~~\overline\psi_k(t,x,y)\to 
 \overline\psi_k(t,-x,y)\gamma^5\gamma^1.
 \label{nn4}
\end{eqnarray}
Due to the  symmetry of the model (\ref{n1}) with respect to each of the discrete 
$\Gamma^5$, $\Gamma^3$ and $\cP$ transformations, different mass 
terms are prohibited to appear in this Lagrangian. Indeed, the most popular mass term has the 
form $m_D\overline \psi_k \psi_k$ (for it we use the notation Dirac mass term), but it breaks 
both, $\Gamma^5$ and $\Gamma^3$, chiral symmetries of the model. There is another well-known 
expression for fermion mass that is often discussed in the literature. This is a mass term of 
the form $m_H\overline \psi_k\tau \psi_k$ (recall, here the 4$\times$4 matrix $\tau$ is defined 
in Appendix \ref{ApC}) and sometimes it is refered to as the Haldane mass term (see, e.g., in Refs. 
\cite{Ebert,Mudry}). \footnote{The appearance of the Haldane mass term is related 
to the parity anomaly in (2+1) dimensions, to generation of the Chern-Simons topological mass 
of gauge fields \cite{Gomes2,Klimenko}, as well as to the integer quantum Hall effect in planar 
condensed matter systems without external magnetic field, etc \cite{Haldane}.} 
But nonzero Haldane mass $m_H$ breaks the parity $\cP$ invariance of the model (although it is 
chirally $\Gamma^5$ and $\Gamma^3$ symmetric). There exist two another mass terms, 
$i m_3\overline \psi_k \gamma^3\psi_k$ and $i m_5\overline \psi_k \gamma^5\psi_k$, the 
dynamical generation of which we are also going to study here. The first one 
breaks $\Gamma^3$ symmetry (but invariant under $\Gamma^5$ and $\cP$), whereas the second 
mass term is not invariant under $\Gamma^5$ and $\cP$ transformations (but it is $\Gamma^3$ 
symmetrical). So the Dirac $m_D$, Haldane $m_H$ and $m_3,m_5$ masses can not be added by hand to the chirally 
$\Gamma^5$, $\Gamma^3$ and parity $\cP$ invariant Lagrangian (\ref{n1}), etc. 

Nevertheless, as can be seen from the analysis of the phase structure 
of the model (\ref{n1}) carried out up to a first order in $G$ within the framework of the CJT
composite operator approach 
(see in the text below), all the above-mentioned masses may be dynamically induced. This means 
that, depending on which behavior mode is chosen for the bare coupling constant $G\equiv 
G(\Lambda)$ vs the cutoff regularization parameter $\Lambda$, the model can dynamically 
generate a phase with one or another nonzero fermion mass. In particular, in order to generate the 
Dirac mass in the model, the bare coupling $G(\Lambda)$ should be positive at sufficiently 
high values of $\Lambda$ and behaves like in Eq. (\ref{440}), whereas to obtain the phase with 
$m_H\ne 0$ the bare coupling $G(\Lambda)$ vs $\Lambda$ should have a negative sign (see in Eq.
(\ref{0480})), etc. 

\subsection{CJT effective action}

Let us define $Z(K)$, the generating functional of the Green's functions of bilocal 
fermion-antifermion composite operators $\sum_{k=1}^N\overline\psi_k^\alpha(x)\psi_{k\beta}(y)$ 
in the framework of a (2+1)D GN model (\ref{n1}) (the corresponding technique 
for theories with four-fermion interaction is elaborated in details, e.g., in Ref. \cite{Rochev}) 
\footnote{Moreover, we should note that in Ref. \cite{Dorey} the CJT composite operator approach 
was applied to (1+1)D GN model.},
\begin{eqnarray}
 Z(K)\equiv\exp(iNW(K))=\int {\cal D}\overline\psi_k {\cal D}\psi_k \exp\Big
(i\Big [ I(\overline\psi,\psi)+\int d^3xd^3y\overline\psi_k^\alpha(x)K_\alpha^\beta(x,y)
\psi_{k\beta}(y) \Big ]\Big ),
 \label{36}
\end{eqnarray}
where $\alpha,\beta =1,2,3,4$ are spinor indices, $K_\alpha^\beta(x,y)$ is a bilocal source of 
the fermion bilinear composite field $\bar\psi_k^\alpha(x)\psi_{k\beta}(y)$ (recall that in 
all expressions the summation over repeated indices is assumed). 
\footnote{We denote a matrix element of an arbitrary matrix (operator) $\hat A$ acting in the 
four dimensional spinor space by the symbol $A^\alpha_\beta$, where the upper (low) index 
$\alpha $($\beta$) is the column (row) number of the matrix $\hat A$. In particular, the matrix 
elements of any $\gamma^\mu$ matrix is denoted by $(\gamma^\mu)^\alpha_\beta$. }
Moreover, $I(\bar\psi,\psi)=\int Ld^3x$, where $L$ is the Lagrangian (\ref{n1}) of a (2+1)-dimensional GN 
model under consideration. It is evident that
$$
I(\overline\psi ,\psi)=\int d^3xd^3y\overline\psi_k^\alpha(x)D_\alpha^\beta(x,y)\psi_{k\beta} (y)+
I_{int}(\overline\psi_k^\alpha\psi_{k\beta}),~~D_\alpha^\beta(x,y)=
\left(\gamma^\nu\right)_\alpha^\beta i\partial_\nu\delta^3(x-y),
$$\vspace{-0.5cm}
\begin{eqnarray}
I_{int}&=& \frac {G}{2N}\int d^3x\left (\bar \psi_k\psi_k\right )^2
=\frac {G}{2N}\int d^3xd^3td^3ud^3v\delta^3 (x-t)\delta^3 (t-u)\delta^3 (u-v) \overline
\psi_k^\alpha(x)\delta_\alpha^\beta\psi_{k\beta}(t) \overline\psi_l^\rho(u)
\delta_\rho^\xi\psi_{l\xi}(v).
 \label{360}
\end{eqnarray}
Note that in Eq. (\ref{360}) and below $\delta^3(x-y)$ (and similar expressions) denotes the 
three-dimensional Dirac delta function. There is an alternative expression for $Z(K)=\exp(iNW(K))$,
\begin{eqnarray}
\exp(iNW(K))&=&\exp\Big (iI_{int}\Big (-i\frac{\delta}{\delta K}\Big )\Big )\int 
{\cal D}\overline\psi_k {\cal D}\psi_k \exp\Big (i
\int d^3xd^3y\overline\psi_k(x)\Big [D(x,y)+K(x,y)\Big ]\psi_k (y)\Big )
\nonumber\\&=&\exp\Big (iI_{int}\Big (-i\frac{\delta}{\delta K}\Big )\Big )\Big 
[\det\big (D(x,y)+K(x,y)\big )\Big ]^N\nonumber\\&=&\exp\Big (iI_{int}\Big 
(-i\frac{\delta}{\delta K}\Big )\Big )\exp \Big [N{\rm Tr}\ln \big (D(x,y)+K(x,y)\big )\Big ],
\label{036}
\end{eqnarray}
where instead of each bilinear form $\bar\psi_k^\alpha(s)\psi_{k\beta}(t)$ appearing in $I_{int}$ of
the Eq. (\ref{360}) we use a variational derivative $-i\delta /\delta K^\beta_\alpha (s,t)$.
Moreover, the Tr-operation in Eq. (\ref{036}) means the trace both over spacetime and spinor coordinates. 
The effective action (or CJT effective action) of the composite bilocal and bispinor operator 
$\bar\psi_k^\alpha(x)\psi_{k\beta}(y)$ is defined as a functional $\Gamma (S)$ of the full 
fermion propagator $S^\alpha_\beta(x,y)$ by a Legendre transformation of the functional 
$W(K)$ entering in Eqs. (\ref{36}) and (\ref{036}),
\begin{eqnarray}
\Gamma (S)=W(K)-\int d^3xd^3y S^\alpha_\beta(x,y)K_\alpha^\beta(y,x),
\label{0360}
\end{eqnarray}
where
\begin{eqnarray}
S^\alpha_\beta(x,y)=\frac{\delta W(K)}{\delta K_\alpha^\beta(y,x)}.
 \label{37}
\end{eqnarray}
Taking into account the relation (\ref{36}), it is clear that $S(x,y)$ is the full fermion 
propagator at $K(x,y)=0$. Hence, in order to construct the CJT effective action $\Gamma (S)$ of Eq. 
(\ref{0360}), it is necessary to solve Eq. (\ref{37}) with respect to $K$ and then to use the obtained expression for $K$ (it is a functional of $S$) in Eq. (\ref{0360}). It is clear from the definition (\ref{0360})-(\ref{37}) that
\begin{eqnarray}
\frac{\delta\Gamma (S)}{\delta S^\alpha_\beta(x,y)}=\int d^3ud^3v\frac{\delta W(K)}{\delta K^\mu_\nu(u,v)}\frac{\delta K^\mu_\nu(u,v)}{\delta S^\alpha_\beta(x,y)}-K_\alpha^\beta(y,x)-\int d^3ud^3v S_\mu^\nu(v,u)\frac{\delta K^\mu_\nu(u,v)}{\delta S^\alpha_\beta(x,y)}.
 \label{037}
\end{eqnarray}
(In Eq. (\ref{037}) and below, the Greek letters $\alpha,\beta,\mu,\nu,$ etc, also denote
the spinor indices, i.e. $\alpha,...\nu,...=1,...,4$.) Now, due to the relation (\ref{37}), it is easy to see that the first term in Eq. (\ref{037}) 
cansels there the last term, so
\begin{eqnarray}
\frac{\delta\Gamma (S)}{\delta S^\alpha_\beta(x,y)}=-K_\alpha^\beta(y,x).
\label{370}
\end{eqnarray}
Hence, in the true GN theory, in which bilocal sources $K_\alpha^\beta(y,x)$ are zero, the full fermion propagator is a solution of the following stationary equation,
\begin{eqnarray}
\frac{\delta\Gamma (S)}{\delta S^\alpha_\beta(x,y)}=0.
\label{0370}
\end{eqnarray}
Note that in the nonperturbative CJT approach the stationary/gap equation (\ref{0370}) 
for fermion propagator $S^\beta_\alpha(x,y)$  is indeed a 
Schwinger--Dyson equation \cite{Rochev}.
Further, in order to simplify the calculations and obtain specific information about 
the phase structure of the model, we calculate the effective action (\ref{0360}) up to a 
first order in the coupling $G$. In this case (see in Appendix \ref{ApB})
\begin{eqnarray}
\Gamma (S)&=&-i{\rm Tr}\ln \big (-iS^{-1}\big )+\int d^3xd^3y S^\alpha_\beta(x,y)D_\alpha^\beta(y,x)\nonumber\\
&+&\frac{G}2\int d^3x \Big [{\rm tr}S(x,x)\Big ]^2 -\frac{G}{2N}\int d^3x~ {\rm tr}\Big [S(x,x)S(x,x)\Big ].
 \label{n420}
\end{eqnarray}
Notice that in Eq. (\ref{n420}) the symbol tr means the trace of an operator 
over spinor indices only, but Tr is the trace operation both over spacetime coordinates and spinor 
indices. Moreover, there the operator $D(x,y)$ is introduced in Eq. (\ref{360}). The stationary 
equation (\ref{0370}) for the CJT effective action (\ref{n420}) looks like 
\begin{eqnarray}
0&=&i\Big [S^{-1}\Big ]^\beta_\alpha(x,y)+D_\alpha^\beta(x,y)+G\delta^\beta_\alpha\delta^3
(x-y)~{\rm tr}S(x,y)-\frac GN S^\beta_\alpha(x,y)\delta^3 (x-y).
\label{n0420}
\end{eqnarray}
Now suppose that $S(x,y)$ is a translationary invariant operator. Then 
\begin{eqnarray}
S^\beta_\alpha(x,y)\equiv S^\beta_\alpha(z)&=&\int\frac{d^3p}{(2\pi)^3}
\overline{S^\beta_\alpha}(p)e^{-ipz},~~~\overline{S^\beta_\alpha}(p)=\int d^3z S^\beta_\alpha(z)e^{ipz},\nonumber\\
\Big (S^{-1}\Big )^\beta_\alpha(x,y)&\equiv& \Big (S^{-1}\Big )^\beta_\alpha(z)=\int\frac{d^3p}{(2\pi)^3}
\overline{(S^{-1})^\beta_\alpha}(p)e^{-ipz},
\label{n43}
\end{eqnarray}
where $z=x-y$ and $\overline{S^\beta_\alpha}(p)$ is a Fourier transformation of $ S^\beta_\alpha(z)$. After Fourier transformation the Eq. (\ref{n0420}) takes the form
\begin{eqnarray}
\overline{(S^{-1})^\beta_\alpha}(p)-i p_\nu(\gamma^\nu)^\beta_\alpha=iG\delta^\beta_\alpha\int\frac{d^3q}{(2\pi)^3}~{\rm tr}\overline{S}(q)
-i\frac GN \int\frac{d^3q}{(2\pi)^3}
\overline{S^\beta_\alpha}(q).
 \label{n043}
\end{eqnarray}
It is clear from Eq. (\ref{n043}) that in the framework of the four-fermion model (\ref{n1}) 
the Schwinger-Dyson equation for fermion propagator $\overline{S}(p)$ reads in the first order
in $G$ like the Hartree-Fock equation for its self-energy operator $\Sigma(p)$. In particular, 
the first and second terms on the right-hand side of Eq. (\ref{n043}) are, respectively, the 
so-called Hartree and Fock contributions to the fermion self energy (for details, see, e.g., 
the section 4.3.1 in Ref. \cite{Buballa}). 

Finally note that both the CJT effective action (\ref{n420}) and its stationary equation 
(\ref{n0420})-(\ref{n043}), in which $G$ is a bare coupling constant, contain ultraviolet 
divergences and need to be renormalized. In the next sections, using different ansatzes for 
propagator $\overline{S}(p)$, we find the corresponding modes of the coupling constant 
$G\equiv G(\Lambda)$ behavior, such that  there occurs a renormalization of the gap equation 
(\ref{n043}) and it is possible to obtain its final solution in the limit $\Lambda\to\infty$.

\section{Possibility for dynamical generation of the Dirac mass }

As it is noted in the previous sections, in three spacetime dimensions there are several 
alternatives for choosing a fermionic mass term of the Lagrangian when fermions are 
transformed, as in the present consideration, according to a reducible four-component spinor 
representation of the Lorentz group (see, e.g., in Appendix \ref{ApC}). One of them is the 
so-called Dirac mass term $m_D\overline\psi_k\psi_k$. Recall that it is $\cP$ 
invariant but breaks chiral invariances $\Gamma^5$ and $\Gamma^3$. Therefore, Dirac mass term
can arise in the model (\ref{n1}) only nonperturbatively, i.e. within the framework of one 
or another nonperturbative approach. And in the present section we study just this possibility 
in the framework of the nonperturbative CJT approach to the model (\ref{n1}). So we seek the 
solution $\overline {S}(p)$ (which is some matrix in the 4-dimensional spinor space) of the 
Eq. (\ref{n043}) in the form
\begin{eqnarray}
\overline{S^{-1}}(p)=i(\hat p+m_D), ~~~{\rm i.e.}~~~\overline {S}(p)=-i
\frac{\hat p-m_D}{p^2-m_D^2},
 \label{430}
\end{eqnarray}
where $p^2=p_0^2-p_1^2-p^2_2$, $\hat p\equiv p_\nu\gamma^\nu$ and the Dirac mass $m_D$ is some 
unknown quantity. Substituting the expression (\ref{430}) into Eq. (\ref{n043}), we obtain for 
$m_D$ the following gap equation
\begin{eqnarray}
m_D=im_D\left (4G-\frac {G}{N}\right )\int\frac{d^3p}{(2\pi)^3}\frac{1}{p^2-m_D^2},
 \label{0430}
\end{eqnarray}
where we took into acount that ${\rm tr}~\hat p=0$, ${\rm tr}~m_D=4m_D$ and 
$\int\frac{d^3p}{(2\pi)^3}\frac{\hat p}{p^2-m_D^2}=0$. After a Wick rotation to Euclidean 
energy-momentum in Eq. (\ref{0430}), i.e., $p_0\to i p_0$, we see that $m_D$ should obey the equation (in which $p^2=p_0^2+p_1^2+p^2_2$)
\begin{eqnarray}
\frac {m_D}G=m_D\left (4-\frac {1}{N}\right )\int\frac{d^3p}{(2\pi)^3}\frac{1}{p^2+m_D^2}.
 \label{44}
\end{eqnarray}
The gap equation (\ref{44}) contains an ultraviolet (UV) divergent integral, i.e. we have to 
first of all regularize this equation. It can be done by using there the spherical coordinate 
system when $\int d^3pf(\sqrt{p_0^2+p_1^2+p_2^2})=4\pi\int_0^\infty p^2dpf(p)$ and 
$p=\sqrt{p_0^2+p_1^2+p_2^2}$. Then, cutting the obtained one-dimensional UV-divergent 
integral by $\Lambda$, we have for $m_D$ the {\it regularized} gap equation
\begin{eqnarray}
\frac {m_D}G=m_D\frac{4N-1}{2N\pi^2}\int_0^\Lambda\frac{p^2}{p^2+m_D^2}dp=
m_D\frac{4N-1}{2N\pi^2}\Big (\Lambda-|m_D|\frac \pi 2+m_D{\cal O}\Big (\frac {m_D}\Lambda\Big )\Big ).
\label{044}
\end{eqnarray}
The gap equation (\ref{044}) is an UV divergent at $\Lambda\to\infty$. This UV divergence can 
be removed from the CJT effective action (\ref{n420}) itself (and in particular from the gap 
equation (\ref{044})), if one demand (which is clear from the form of Eq. (\ref{044})) that 
the bare coupling constant $G\equiv G(\Lambda)$ has the following dependence on the 
cutoff parameter $\Lambda$,
\begin{eqnarray}
\frac 1{G(\Lambda)}=\frac{4N-1}{2N\pi^2}\Big (\Lambda+g_D\frac \pi 2+g_D{\cal O}\Big (\frac {g_D}\Lambda\Big )\Big ),
\label{440}
\end{eqnarray}
where $g_D$ is a finite $\Lambda$-independent and renormalization group invariant quantity with 
dimension of mass. Instead of the bare coupling constant $G$, the parameter $g_D$ can be 
considered as a new free parameter of the model. Now, substituting Eq. (\ref{440}) into Eq. (\ref{044}) we obtain 
in the limit $\Lambda\to\infty$ a finite, i.e. {\it renormalized}, equation for Dirac mass $m_D$,
\begin{eqnarray}
m_D(g_D+|m_D|)=0 \label{404}
\end{eqnarray}
from which it is clear that at $g_D>0$ only a trivial solution of the gap equation (\ref{44})-(\ref{404})
exists, 
$m_D=0$, i.e. the dynamical generation of the fermion mass is impossible. However, at $g_D<0$ 
there are two solutions, (i) $m_D=0$ and (ii) $m_D=-g_D$, of the 
gap equation (\ref{404}). To find which of the solutions of the gap equation is more preferable in this 
case, it is necessary to consider the so-called CJT effective potential $V(S)$ of the model 
which is defined on the basis of the CJT effective action (\ref{n420}) by the following 
relation \cite{CJT,Casalbuoni}
\begin{eqnarray}
V(S)\int d^3x\equiv -\Gamma(S)\Big |_{\rm transl.-inv.~S(x,y) },
 \label{470}
\end{eqnarray}
where $S(x,y)$ is a translation invariant quantity, i.e. $S(x,y)\equiv S(x-y)$, as assumed
in Eqs. (\ref{n43}) and (\ref{430}). It is evident that for arbitrary values of the bare coupling
constant $G$ the CJT effective potential (\ref{470}) is UV-divergent unrenormalized quantity.
However, if $G$ is constrained by the condition (\ref{440}), then all UV divergences of $V(S)$ are 
eliminated, and for fermion propagator of the form (\ref{430}) it looks like
\begin{eqnarray}
V(S)\equiv V(m_D)=\frac{1}{6\pi}\left (2|m_D|^3+3g_Dm_D^2\right )
\label{471}
\end{eqnarray}
(notice that this expression is valid up to unessential $m_D$-independent infinite constant). Hence, we see 
that at $g_D<0$ $V(m_D=0)=0$ and it is larger than $V(m_D=-g_D)=-|g_D|^3/(6\pi)$. 
So one can conclude that if bare coupling constant $G(\Lambda)$ behaves vs $\Lambda$ as in Eq. 
(\ref{440}),
then at $g_D<0$ a Dirac mass $m_D$ equal to $(-g_D)$ is generated dynamically in the system and for this particular 
behavior of $G(\Lambda)$ vs $\Lambda$ the phase with spontaneous breaking of chiral 
symmetries (\ref{n4}) is realized in the system. 

Note that the phase structure of the GN model (\ref{n1}) can also be investigated within the 
framework of the large-$N$ expansion technique, where it is supposed that at $N\to\infty$ the 
bare coupling constant $G\to$ const $\ne 0$. And it was shown in a lot of papers (see, e.g.,
in Refs. \cite{Rosenstein,GN,Modugno}) that in this case 
the dynamical generation of the Dirac mass and spontaneous breaking of the discrete chiral
symmetries (\ref{n4}) takes place in the leading order of the large-$N$ expansion. 
In the present section we have shown that qualitatively the same result is obtained in the 
first order in $G$ within the CJT composite operator approach, if coupling constant is 
constrained by the condition (\ref{440}). Since in this case $G\to {\rm const}$ at $N\to\infty$,
we may conclude that the CJT composite operator approach reproduces the results of the 
large-$N$ expansion technique. 

In conclusion, about one more property of the GN model (\ref{n1}), considered in the 
CJT effective action approach. It follows from the condition (\ref{440}), 
which leads to renormalizability of the model \footnote{At least in the first order in $G$ 
of the nonperturbative effective action approach for composite operators.}, as well as to dynamical generation of a nonzero
Dirac mass and the appearance of a phase with spontaneous breaking of chiral symmetries 
$\Gamma^5$ and $\Gamma^3$ (\ref{n4}).
Using relation (\ref{440}), it can be shown that there exists a nonzero UV-stable fixed point 
$\lambda_D$ in the model. It is associated with a so-called Callan-Simanzik $\beta (\lambda) $-function defined by 
the relation (see, e.g., in Sec. 2.7 of Ref. \cite{Rosenstein}) 
\begin{eqnarray}
\beta (\lambda)=\Lambda\partial\lambda/\partial\Lambda,
 \label{473}
\end{eqnarray}
where $\lambda\equiv\Lambda G(\Lambda)$ is the dimensionless bare coupling of the model 
(\ref{n1}). Using in Eq. (\ref{473}) the relation (\ref{440}) for $G(\Lambda)$, it is possible 
to obtain
\begin{eqnarray}
\beta (\lambda)=\frac{\lambda}{\lambda_D}(\lambda_D-\lambda),
 \label{474}
\end{eqnarray}
where $\lambda_D=\frac{2N\pi^2}{4N-1}$ is a zero of the $\beta (\lambda)$ function, and it is 
an UV-stable fixed point of the model. It means that in the continuum limit (large $\Lambda$) 
the dimensionless bare coupling $\lambda (\Lambda)$ approaches ultraviolet fixed point 
$\lambda_D$. This feature of the coupling constant $\lambda (\Lambda)$ can be easily seen 
directly from Eq. (\ref{440}) at $\Lambda\to\infty$. Moreover, it is also clear from Eq. (\ref{440})
that at rather large values of $\Lambda$ the relation $\lambda (\Lambda)-\lambda_D 
\sim -g_D/\Lambda$ is valid. As a result, we see that at $\lambda >\lambda_D$ the chiral 
symmetry breaking phase is realized (since in 
this case $g_D$ must be negative), but at rather small values of the dimensionless bare coupling 
$\lambda$ ($<\lambda_D$) the symmetry of the model remains intact, since in this case the 
relation $g_D>0$ must be fulfilled.

\section{Possibility for dynamical generation  of the Haldane mass }

Let us now explore the possibility that the solution of the gap equation (\ref{n043}) has the form 
\begin{eqnarray}
\overline{S^{-1}}(p)=i(\hat p+\tau m_H), ~~~{\rm i.e.}~~~\overline {S}(p)=-i\frac{\hat p-\tau m_H}{p^2-m_H^2},
 \label{0470}
\end{eqnarray}
where $\tau$ is the 4$\times$4 matrix presented by Eq. (\ref{C6}) (see in Appendix \ref{ApC}) and 
other notations are the same as in the formula (\ref{430}). In this case the so-called 
Haldane mass $m_H$ could arise dynamically and, as a result, the parity $\cP$-odd phase 
could be realized in the system (but chiral symmetries (\ref{n4}) are not broken spontaneously in this 
phase). Substituting Eq. (\ref{0470}) into Eq. (\ref{n043}), we obtain for $m_H$ the following 
gap equation
\begin{eqnarray}
m_H=-\frac {im_HG}{N}\int\frac{d^3p}{(2\pi)^3}\frac{1}{p^2-m_H^2},
 \label{48}
\end{eqnarray}
where we took into acount that ${\rm tr}~\hat p=0$, ${\rm tr}~\tau=0$ and 
$\int\frac{d^3p}{(2\pi)^3}\frac{\hat p}{p^2-m^2}=0$. After a Wick rotation in Eq. (\ref{48})
to Euclidean energy-momentum, i.e. $p_0\to i p_0$, we see that $m_H$ should obey the equation (in which $p^2=p_0^2+p_1^2+p^2_2$)
\begin{eqnarray}
\frac {m_H}G=-\frac {m_H}{N}\int\frac{d^3p}{(2\pi)^3}\frac{1}{p^2+m_H^2},
\label{048}
\end{eqnarray}
Using in the three-dimensional integral of Eq. (\ref{048}) the spherical coordinate system 
(see the remark just after Eq. (\ref{44})), it is possible to reduce it to a one-dimensional 
UV-divergent intergral. Cutting it by $\Lambda$,  we have for $m_H$ the following {\it regularized}
gap equation
\begin{eqnarray}
\frac {m_H}G=\frac{-m_H}{2N\pi^2}\int_0^\Lambda\frac{p^2}{p^2+m_H^2}dp=\frac{-m_H}{2N\pi^2}\Big (\Lambda-|m_H|\frac \pi 2+m_H{\cal O}\Big (\frac {m_H}\Lambda\Big )\Big ).
 \label{480}
\end{eqnarray}
The UV divergence can be removed from the gap equation (\ref{480}) if we require 
 (it is clear from the form of this equation) the following behavior of the bare coupling 
 constant $G\equiv G(\Lambda)$ vs $\Lambda$,
\begin{eqnarray}
\frac 1{G(\Lambda)}=\frac{-1}{2N\pi^2}\Big (\Lambda+g_H\frac \pi 2+g_H{\cal O}\Big (\frac {g_H}\Lambda\Big )\Big ),
\label{0480}
\end{eqnarray}
where $g_H$ is a finite $\Lambda$-independent and renormalization group invariant quantity. 
It can be considered as a new free parameter of the model. Now, comparing the Eqs. (\ref{0480})
and (\ref{480}), we obtain in the limit $\Lambda\to\infty$ for the Haldane mass $m_H$ the 
following {\it renormalized}, i.e. without UV divergences, gap equation
\begin{eqnarray}
m_H(g_H+|m_H|)=0. \label{4040}
\end{eqnarray}
Hence, at $g_H>0$ only a trivial solution of the gap equation (\ref{4040}) exists, $m_H=0$,
and symmetry of the model remains intact. 
However, at $g_H<0$ there are two solutions, (i) $m_H=0$ and (ii) $m_H=-g_H$, of this
gap equation. To find which of the solutions of the gap equation is more preferable in this 
case, it is necessary to compare the corresponding values of the CJT effective potential 
(\ref{470}). Using in the formula (\ref{n420}) for $\Gamma(S)$ the expressions (\ref{0470}) and 
(\ref{0480}) for fermion propagator $S(x-y)$ and bare coupling constant $G$, respectively, one can 
obtain in this case, due to the relation (\ref{470}), the following CJT effective potential at $\Lambda\to\infty$ 
(up to unessential $m_H$-independent infinite constant)
\begin{eqnarray}
V(S)\equiv V(m_H)=\frac{1}{6\pi}\left (2|m_H|^3+3g_Hm_H^2\right ).
 \label{472}
\end{eqnarray} 
Then, it is clear from Eq. (\ref{472}) that at $g_H<0$ $V(m_H=-g_H)=-|g_H|^3/(6\pi)$, and this quantity is smaller 
than $V(m_H=0)=0$. This allows us to conclude that if in the original model (\ref{n1}) the 
bare coupling constant $G$ behaves vs $\Lambda$ like expression (\ref{0480}) and $g_H<0$, 
then the system undergoes a dynamic generation of the Haldane mass, i.e. a phase with spontaneous
violation of parity $\cP$ is realized.  

Let us now discuss the possibility of dynamically generating this phase in terms of the 
dimensionless bare coupling constant $\lambda\equiv\lambda(\Lambda)=\Lambda G(\Lambda)$, where 
$G(\Lambda)$ is given by Eq. (\ref{0480}). It is clear that in this case for a sufficiently 
high values of $\Lambda\gg |g_H|$ both the dimensional bare coupling $G (\Lambda)$ and
the dimensionless coupling $\lambda$ are negative. In addition, it easy to see
that at $\Lambda\to\infty$ the dimensionless bare coupling $\lambda$ tends to
the quantity $\lambda_H=-2N\pi^2$, which is called the UV-stable fixed point of the model, and 
it is the zero of the Callan-Simanzik $\beta(\lambda)=\Lambda\partial\lambda/\partial
\Lambda$ function. (Notice that in the case under consideration, the $\beta(\lambda)$-function 
has the same form as one from Eq. (\ref{474}), in which $\lambda_D$ must be replaced by 
$\lambda_H$.) Then the relation 
\begin{eqnarray}
\lambda (\Lambda)-\lambda_H 
\sim \frac{2\pi^2Ng_H}{\Lambda}
 \label{475}
\end{eqnarray} 
can be obtained. It follows from Eq. (\ref{475}) that on the negative $\lambda$-semiaxis the 
UV-fixed point $\lambda_H$ separates the symmetric phase from the one where the parity $\cP$ 
is spontaneously broken. Indeed, if $\lambda<\lambda_H$ then, as it is clear from Eq. (\ref{475}), 
$g_H$ must be negative, which corresponds to $\cP$-odd phase with dynamical generation
of the Haldane mass $m_H$, while at 
$\lambda>\lambda_H$ we have $g_H>0$ and symmetric phase of the model (see the text below Eq.
(\ref{4040})). 

Since $\lambda_H\to -\infty$ at $N\to\infty$, we may conclude that 
in the limit of large $N$ the (2+1)-D GN model (\ref{n1}) cannot have a $\cP$-odd phase and 
Haldane mass  cannot arise dynamically, i.e. in this limit for arbitrary negative values $\lambda$
only symmetric phase can be realized. In other words, the generation of the Haldane mass in 
the model (\ref{n1}) is the effect which cannot be observed in the leading order of the 
$1/N$-expansion technique. The similar result
was obtained in Ref. \cite{Appelquist2} where it was proved in the framework of (2+1)-D 
quantum electrodynamics that using the large-$N$ expansion method it is not possible to observe
spontaneous parity breaking as well as the Haldane mass generation. 

\section{Possibility for dynamical generation  of the $m_5$ and $m_3$  mass terms }

Finally, let us explore the possibility that the solution of the gap equation (\ref{n043}) has the form 
\begin{eqnarray}
\overline{S^{-1}}(p)=i(\hat p+i\gamma^5 m_5+i\gamma^3 m_3), ~~~{\rm i.e.}~~~
\overline {S}(p)=-i\frac{\hat p+i\gamma^5 m_5+i\gamma^3 m_3}{p^2-(m_3^2+m_5^2)}.
 \label{0471}
\end{eqnarray}
It corresponds to a dynamically generated mass term of the form ${\cal M}=\left (m_5\overline
\psi_k i\gamma^5\psi_k
+m_3\overline\psi_k i\gamma^3\psi_k\right )$ in the Lagrangian (\ref{n1}) (the Hermitian matrices $\gamma^{3,5}$
are presented in Appendix \ref{ApC}). Since we suppose that $m_5$ and $m_3$ are some real 
numbers, this mass term is a Hermitian one. And it is not invariant under each of the 
discrete transformations (\ref{n4}) or (\ref{nn4}) (at nonzero $m_3$ and $m_5$).
Substituting Eq. (\ref{0471}) into Eq. (\ref{n043}) and taking into account the technical 
details discussed in previous two sections, one can obtain for $m_3$ and $m_5$ 
the following system of gap equations
\begin{eqnarray}
m_3&=&\frac {im_3G}{N}\int\frac{d^3p}{(2\pi)^3}\frac{1}{p^2-(m_3^2+m_5^2)},\nonumber\\
 m_5&=&\frac {im_5G}{N}\int\frac{d^3p}{(2\pi)^3}\frac{1}{p^2-(m_3^2+m_5^2)}.\label{49}
\end{eqnarray}
 After a Wick rotation in Eq. (\ref{49})
to Euclidean energy-momentum, i.e. $p_0\to i p_0$, we see that $(m_3,m_5)$ should obey the 
equation system (in which $p^2=p_0^2+p_1^2+p^2_2$)
\begin{eqnarray}
\frac{m_3}{G}&=&\frac {m_3}{N}\int\frac{d^3p}{(2\pi)^3}\frac{1}{p^2+(m_3^2+m_5^2)},\nonumber\\
\frac{m_5}{G}&=&\frac {m_5}{N}\int\frac{d^3p}{(2\pi)^3}\frac{1}{p^2+(m_3^2+m_5^2)}.\label{049}
\end{eqnarray}
This system of equations contains UV-divergent integrals, i.e. it is unrenormalized. For its 
regularization, we use, as in the two previous sections, the spherical coordinate system and 
reduce the three-dimensional integral of Eqs. (\ref{049}) to one-dimensional 
UV-divergent intergral (see the remark just after Eq. (\ref{44})). Cutting off the region of 
integration in it by $\Lambda$,  we have for $(m_3,m_5)$ the following {\it regularized} gap 
equations
\begin{eqnarray}
\frac {m_3}G&=&\frac{m_3}{2N\pi^2}\int_0^\Lambda\frac{p^2}{p^2+m_3^2+m_5^2}dp,\nonumber\\
 \frac {m_5}G&=&\frac{m_5}{2N\pi^2}\int_0^\Lambda\frac{p^2}{p^2+m_3^2+m_5^2}dp.
 \label{490}
\end{eqnarray}
Notice that at $\Lambda\to\infty$ an integral term in Eqs.  (\ref{490}) has the 
following asymptotic expansion
\begin{eqnarray}
\int_0^\Lambda\frac{p^2}{p^2+m_3^2+m_5^2}dp=\Lambda-\frac \pi 2\sqrt{m_3^2+m_5^2}+
\sqrt{m_3^2+m_5^2}{\cal O}\left (\frac {\sqrt{m_3^2+m_5^2}}\Lambda\right ).
 \label{4900}
\end{eqnarray}
Hence, taking into account the expansion (\ref{4900}), the UV divergence can be removed from 
the gap equations (\ref{490}) if we require 
 (it is clear from the form of this equation system) the following behavior of the bare coupling 
 constant $G\equiv G(\Lambda)$ vs $\Lambda$,
\begin{eqnarray}
\frac 1{G(\Lambda)}=\frac{1}{2N\pi^2}\Big (\Lambda+g\frac \pi 2+g{\cal O}\Big (\frac {g}
\Lambda\Big )\Big ),
\label{0490}
\end{eqnarray}
where $g$ is a finite $\Lambda$-independent and renormalization group invariant quantity, and 
it can also be considered as a new free parameter of the model. Now, comparing Eqs. 
(\ref{0490})
and (\ref{490}), we obtain in the limit $\Lambda\to\infty$ the 
following renormalized, i.e. without UV divergences, gap equations for the masses $m_3$ and $m_5$
\begin{eqnarray}
m_3\left (g+\sqrt{m_3^2+m_5^2}\right )&=&0,\nonumber\\
m_5\left (g+\sqrt{m_3^2+m_5^2}\right )&=&0.
\label{40400}
\end{eqnarray}
Hence, at $g>0$ only a trivial solution of the gap equations (\ref{40400}) exists, $m_3=m_5=0$,
and symmetry of the model remains intact. 
However, at $g<0$ there are two solutions, (i) $m_3=0, m_5=0$ and (ii)
$m_3=|g|\cos\alpha, m_5=|g|\sin\alpha$ (where $0\le \alpha\le\pi/2$ is some arbitrary fixed 
angle), of the system (\ref{40400}) of gap equations. 
To find which of the solutions, (i) or (ii), is more preferable in this 
case, it is necessary to compare the corresponding values of the CJT effective potential $V(S)$
(\ref{470}). It can be easily found if in the expression (\ref{n420}) for $\Gamma(S)$ we use 
the propagator $S(x-y)$ and the bare coupling constant $G(\Lambda)$, given by formulas 
(\ref{0471}) and (\ref{0490}), respectively. After a series of simple calculations, 
we obtain in this case the following CJT effective potential at $\Lambda\to\infty$ 
(up to unessential $m_3$- and $m_5$-independent infinite constant)
\begin{eqnarray}
V(S)\equiv V(m_3,m_5)=\frac{1}{6\pi}\left (2(m_3^2+m_5^2)^{3/2}+3g (m_3^2+m_5^2)\right ).
 \label{492}
\end{eqnarray} 
It is clear from Eq. (\ref{492}) that at $g<0$ effective potential takes on the solution (ii) 
the value $-|g|^3/(6\pi)$, and this quantity is smaller than $V(m_3=0,m_5=0)=0$. This allows us 
to conclude that if in the original model (\ref{n1}) the 
bare coupling constant $G$ behaves vs $\Lambda$ like in the expression (\ref{0490}) with $g<0$,
then the system undergoes a dynamic generation of the $m_3=|g|\cos\alpha$ and $m_5=|g|\sin\alpha$
masses, i.e. a phase with spontaneous violation of all discrete symmetries (\ref{n4}) and 
(\ref{nn4}) is realized in the model (if $\alpha\ne 0,\pi/2$). But if $\alpha =0$ then only 
$m_3$ mass is generated and $\Gamma^3$ chiral symmetry is dynamically violated. However, at 
$\alpha =\pi/2$ only $m_5$ mass appears dynamically and in this phase both chiral $\Gamma^5$ 
and parity $\cP$ are broken spontaneously. Finally notice that at $g<0$ in all above mentioned 
cases, i.e. at arbitrary values of angle parameter $\alpha$, the genuine physical fermion 
mass, which is indeed a pole of the 
fermion propagator (\ref{0471}), is equal to $\sqrt{m_3^2+m_5^2}=|g|$.

In terms of dimensionless bare coupling constant $\lambda\equiv\lambda(\Lambda)=
\Lambda G(\Lambda)$, where $G(\Lambda)$ is given by Eq. (\ref{0490}), the situation looks as 
follows. It is clear that in this case for a sufficiently 
high values of $\Lambda\gg |g|$ both the dimensional bare coupling $G (\Lambda)$ and
the dimensionless coupling $\lambda$ are positive. In addition, it easy to see
that at $\Lambda\to\infty$ the dimensionless bare coupling $\lambda$ tends to
the quantity $\lambda_{35}\equiv 2N\pi^2$, which is the UV-stable fixed point of the model, 
and 
it is the zero of the Callan-Simanzik $\beta(\lambda)=\Lambda\partial\lambda/\partial
\Lambda$ function. (Notice that in the case under consideration, i.e. when $G(\Lambda)$ is 
defined by Eq. (\ref{0490}), the $\beta(\lambda)$-function 
has the same form as one from Eq. (\ref{474}), in which $\lambda_D$ must be replaced by 
$\lambda_{35}$.) Then the relation 
\begin{eqnarray}
\lambda -\lambda_{35} \sim -\frac{2\pi^2Ng}{\Lambda}
 \label{495}
\end{eqnarray} 
can be obtained. It follows from Eq. (\ref{495}) that on the positive $\lambda$-semiaxis the 
UV-fixed point $\lambda_{35}$ separates the symmetric phase from the one in which fermions 
are massive.  Indeed, if $\lambda>\lambda_{35}$ then, as it is clear from Eq. 
(\ref{495}), the parameter 
$g$ must be negative, which corresponds to a dynamical generation of the mass term 
${\cal M}=\left (m_5\overline\psi i\gamma^5\psi +m_3\overline\psi i\gamma^3\psi\right )$
in the Lagrangian (which indeed corresponds to a physical fermion mass equal to $|g|$), while at 
$\lambda<\lambda_{35}$ we have $g>0$ from Eq. (\ref{495}) and symmetric phase of the model 
(see the text below Eq. (\ref{40400})). 

\section{Summary and conclusions}

In the present paper we have studied phase structure of the $N$-flavored massless 
(2+1)-dimensional GN model (\ref{n1}), using a bilocal source formalism in order to construct 
the CJT effective action $\Gamma(S)$ (\ref{0360}) for the composite bifermion operator 
$\overline\psi_k (x)\psi_k (y)$. (Of course, in this case the number of fermionic multiplets 
could be fixed from the very beginning, but since we want to compare the results of 
the CJT approach with large-$N$ expansion method, throughout the paper $N$ is a free parameter.) 
In fact, $\Gamma(S)$ is a functional of a full fermion 
propagator $S(x,y)$ (see in the section II), and in this case, in order to find the true 
fermion propagator of the original GN model and to determine what kind of fermionic mass term 
can arise dynamically in the model (\ref{n1}), it is sufficient to solve the corresponding 
stationarity (gap) equation (\ref{0370}) for the functional $\Gamma(S)$. 

Note that in (2+1)-D condensed matter systems up to 36 different order 
parameters bilinear in Fermi fields, or mass terms, can exist \cite{Mudry}. So for simplicity,
in this paper we have considered dynamical emergence of only four of them, Dirac $m_D\overline\psi_k\psi_k$, Haldane $m_H\overline\psi_k\tau\psi_k$ or of 
the form $m_5\overline\psi_k i\gamma^5\psi_k$ and $m_3\overline\psi_k 
i\gamma^3\psi_k$ (see the corresponding definitions in section II A), 
within the framework of the simplest GN model
using the CJT composite operator approach. The appearance of other possible masses can be 
analyzed in a similar way (including models with a more complex four-fermion structure). 

Moreover, the study of the occurrence of each of the above-mentioned masses 
of fermions is carried out, for simplicity, using the CJT effective action and its gap equation calculated up 
to the first order in the coupling constant $G$ (see, respectively, Eqs. (\ref{n420}) and 
(\ref{n0420})). Then, assuming a translational invariance of the fermionic propagator $S(x,y)$
in each of the cases under investigation, we conclude:

(i) Finite renormalized, i.e. without ultraviolet divergences, gap equation (\ref{404}) for 
the Dirac mass $m_D$ arises only when the bare coupling constant $G\equiv G(\Lambda)$ has a 
dependence on the cutoff parameter $\Lambda$ presented by the formula (\ref{440}). In this 
case, only at $g_D<0$ the Dirac mass $m_D$ can arise dynamically and it looks like $m_D=-g_D$, 
where $g_D$ is a finite and renormalization group invariant quantity with dimension of mass. 
At sufficiently high values of the cutoff parameter $\Lambda$, the dimensionless bare coupling 
constant $\lambda\equiv\Lambda G(\Lambda)$ is always a positive quantity such that 
$\lambda\to\lambda_D\equiv\frac{2N\pi^2}{4N-1}$ at $\Lambda\to\infty$, where $\lambda_D$ is 
the so-called UV-stable fixed point of the model. Moreover, in this case only at 
$\lambda>\lambda_D$ the chiral symmetry breaking occurs and fermions acquire dynamically 
the nonzero Dirac mass $m_D$ (at $\lambda<\lambda_D$ we have $m_D\equiv 0$).  As a result, 
we see that if $G(\Lambda)$ vs $\Lambda$ is defined by Eq. (\ref{440}), then the phase 
structure of the (2+1)-D GN model (\ref{n1}), considered in the framework of the CJT effective
action approach for composite operators, is qualitatively the same as 
if we considered it with the help of the large-$N$ expansion technique (see, e.g., in Refs.
\cite{Rosenstein,GN,Modugno}).

(ii) A nontrivial and renormalized, i.e. without UV divergences, gap equation (\ref{4040}) 
for the 
Haldane mass $m_H$ appears in the model only when the bare coupling constant 
$G(\Lambda)$ vs $\Lambda$ is presented by Eq. (\ref{0480}). Under this constraint on $G$, 
a nonzero Haldane mass $m_H$ can arise dynamically only at $g_H<0$, when $m_H=-g_H$, and 
in the model a phase is realized in which spatial parity $\cP$ is broken spontaneously. 
It should be noted that within the framework of the (2+1)-D GN model (\ref{n1}), this phase 
is allowed to exist only at finite values of $N$ when the CJT 
effective action method for composite operators \cite{CJT,Peskin,Casalbuoni,Dorey,Rochev} 
is used.
The phase cannot be noticed in the model (\ref{n1}), for example, in the framework of
a well-known large-$N$ expansion method, etc. In this case, i.e. when $G(\Lambda)$ vs 
$\Lambda$ is defined by Eq. (\ref{0480}), the dimensionless bare coupling 
$\lambda\equiv\Lambda G(\Lambda)$ is negative and $\lambda\to\lambda_H
\equiv -2\pi^2N$ at $\Lambda\to\infty$, where $\lambda_H$ is also an UV-fixed point of the 
model. Note that Haldane mass $m_H$ is generated only at $\lambda<\lambda_H$ 
(so at $N\to\infty$ the effect disappears). And at $\lambda>\lambda_H$ the symmetry of 
the model is not broken spontaneously, i.e. in this case $m_H\equiv 0$.

Note once again that in the regime (\ref{0480}) of the coupling constant $G(\Lambda)$, 
spontaneous violation of parity $\cP$ occurs in the model, and the Hadane mass $m_H$ arises 
dynamically. In this case, if in addition to fermions there are gauge fields in the model 
(\ref{n1}), then due to the nonzero Haldane mass $m_H$, the topological mass of the gauge 
fields is induced, and the so-called Chern-Simons term is induced dynamically in the system 
\cite{Vshivtsev,Gomes2,Klimenko}. In this case such phenomena as 
quantum Hall effect, exotic statistics and fractional spin as well as high-temperature 
superconductivity can be observed (see, e.g., the discussion in Refs. 
\cite{Vshivtsev,Khudyakov}). 

(iii) Finally, in section V we have investigated, using the CJT effective action approach,
the possibility for dynamical generation of 
the mass term ${\cal M}=(m_5\overline\psi_k i\gamma^5\psi_k+m_3\overline\psi_k 
i\gamma^3\psi_k)$ which in fact involves two qualitatively different fermion masses, 
$m_5$ and $m_3$. It turns out that this effect can manifest itself only when the bare coupling
constant $G(\Lambda)$ behaves vs $\Lambda$ as in Eq. (\ref{0490}), i.e. it is a positive quantity. 
In this case both the CJT effective 
action $\Gamma(S)$ and its stationary equations can be renormalized (also in the first order
in $G$), and one can find finite 
expressions for $m_5$ and $m_3$ such that $\sqrt{m_3^2+m_5^2}=|g|$ when $g<0$ (note that $g$ 
appears in Eq. (\ref{0490}) as a renormalization group invariant free parameter of the model). 
In this phase 
all discrete symmetries (\ref{n4}) and (\ref{nn4}) of the model are spontaneously broken down, 
and fermions aquire dynamically a mass $M_F\equiv\sqrt{m_3^2+m_5^2}$ equal to $|g|$ 
(it is a singularity of the fermion 
propagator (\ref{0471})). Note that if $g>0$ then $M_F=0$. In this case, i.e. when 
$G(\Lambda)$ vs $\Lambda$ is defined by Eq. (\ref{0490}), the dimensionless bare coupling 
$\lambda\equiv\Lambda G(\Lambda)$ is positive and $\lambda\to\lambda_{35}
\equiv 2\pi^2N$ at $\Lambda\to\infty$, where $\lambda_{35}$ is also an UV-fixed point of the 
model. The fermion mass $M_F$ is generated only at $\lambda>\lambda_{35}$ 
(so at $N\to\infty$ the effect disappears, it cannot be observed, e.g., in the framework of 
$1/N$ technique). And at $\lambda<\lambda_{35}$ the symmetry of 
the model is not broken spontaneously, i.e. in this case $M_F\equiv 0$.

As a result, we see that three different phases, discribed above in the items (i), (ii) and 
(iii), can be observed in the simplest (2+1)-D GN model (\ref{n1}) with the help of 
nonperturbative CJT effective action approach for composite operators. One of them, the phase 
(i), is the same that can be detected using the large-$N$ method. Two others, the phases 
(ii) and (iii),  exist only at finite $N$ and cannot be observed by $1/N$ expansion technique.
Note also that each of the phases is characterized by its own behavior of the bare coupling
constant $G(\Lambda)$ vs $\Lambda$.

Finally, let us pay attention to another important feature of our study. It is well known that (2 + 1)-D 
GN model (\ref{n1}) is a nonrenormalizable from the point of view of the ordinary perturbation 
theory. But within the framework of the large-$N$ expansion technique, it is renormalizable 
since only three counterterms need to be introduced into the model in order to eliminate 
all UV divergences in each order in $1/N$ \cite{Rosenstein}. To investigate the phase structure of 
this model, we used another nonperturbative approach, the CJT composite operator method \cite{CJT}, and within this approach we renormalized the model in the first order in $G$. However, strictly speaking, it remains unclear whether the model is renormalizable (whether this could be done
in higher orders in $G$) in the framework of the CJT approach. And it can be considered as a 
subject of future studies.

\section{ACKNOWLEDGMENTS}

R.N.Z. is grateful for support of the Foundation for the Advancement of Theoretical Physics and 
Mathematics BASIS.

\appendix
\section{Algebra of the $\gamma$ matrices in the case of SO(2,1) group}
\label{ApC}

The two-dimensional irreducible representation of the (2+1)-dimensional
Lorentz group SO(2,1) is realized by the following $2\times 2$
$\tilde\gamma$-matrices:
\begin{eqnarray}
\tilde\gamma^0=\sigma_3=
\left (\begin{array}{cc}
1 & 0\\
0 &-1
\end{array}\right ),\,\,
\tilde\gamma^1=i\sigma_1=
\left (\begin{array}{cc}
0 & i\\
i &0
\end{array}\right ),\,\,
\tilde\gamma^2=i\sigma_2=
\left (\begin{array}{cc}
0 & 1\\
-1 &0
\end{array}\right ),
\label{C1}
\end{eqnarray}
acting on two-component Dirac spinors. They have the properties:
\begin{eqnarray}
Tr(\tilde\gamma^{\mu}\tilde\gamma^{\nu})=2g^{\mu\nu};~~
[\tilde\gamma^{\mu},\tilde\gamma^{\nu}]=-2i\varepsilon^{\mu\nu\alpha}
\tilde\gamma_{\alpha};~
~\tilde\gamma^{\mu}\tilde\gamma^{\nu}=-i\varepsilon^{\mu\nu\alpha}
\tilde\gamma_{\alpha}+g^{\mu\nu},
\label{C2}
\end{eqnarray}
where $g^{\mu\nu}=g_{\mu\nu}=diag(1,-1,-1),
~\tilde\gamma_{\alpha}=g_{\alpha\beta}\tilde\gamma^{\beta},~
\varepsilon^{012}=1$.
There is also the relation:
\begin{eqnarray}
Tr(\tilde\gamma^{\mu}\tilde\gamma^{\nu}\tilde\gamma^{\alpha})=
-2i\varepsilon^{\mu\nu\alpha}.
\label{C3}
\end{eqnarray}
Note that the definition of chiral symmetry is slightly unusual in
(2+1)-dimensions (spin is here a pseudoscalar rather than a (axial)
vector). The formal reason is simply that there exists no other $2\times 2$ matrix anticommuting with the Dirac matrices $\tilde\gamma^{\nu}$
which would allow the introduction of a $\gamma^5$-matrix in the
irreducible representation. The important concept of 'chiral'
symmetries  and their breakdown by mass terms can nevertheless be
realized also in the framework of (2+1)-dimensional quantum field
theories by considering a four-component reducible representation
for Dirac fields. In this case the Dirac spinors $\psi$ have the
following form:
\begin{eqnarray}
\psi(x)=
\left (\begin{array}{cc}
\tilde\psi_{1}(x)\\
\tilde\psi_{2}(x)
\end{array}\right ),
\label{C4}
\end{eqnarray}
with $\tilde\psi_1,\tilde\psi_2$ being two-component spinors.
In the reducible four-dimensional spinor representation one deals
with 4$\times$4 $\gamma$-matrices:
$\gamma^\mu=diag(\tilde\gamma^\mu,-\tilde\gamma^\mu)$, where
$\tilde\gamma^\mu$ are given in (\ref{C1}) (This particular reducible representation for 
$\gamma$-matrices is used, e.g., in Ref. \cite{Appelquist}). One can easily show, that
($\mu,\nu=0,1,2$):
\begin{eqnarray}
&&Tr(\gamma^\mu\gamma^\nu)=4g^{\mu\nu};~~
\gamma^\mu\gamma^\nu=\sigma^{\mu\nu}+g^{\mu\nu};~~\nonumber\\
&&\sigma^{\mu\nu}=\frac{1}{2}[\gamma^\mu,\gamma^\nu]
=diag(-i\varepsilon^{\mu\nu\alpha}\tilde\gamma_\alpha,
-i\varepsilon^{\mu\nu\alpha}\tilde\gamma_\alpha).
\label{C5}
\end{eqnarray}
In addition to the  Dirac matrices $\gamma^\mu~~(\mu=0,1,2)$ there
exist two other matrices, $\gamma^3$ and $\gamma^5$, which anticommute
with all $\gamma^\mu~~(\mu=0,1,2)$ and with themselves
\begin{eqnarray}
\gamma^3=
\left (\begin{array}{cc}
0~,& I\\
I~,& 0
\end{array}\right ),\,
\gamma^5=\gamma^0\gamma^1\gamma^2\gamma^3=
i\left (\begin{array}{cc}
0~,& -I\\
I~,& 0
\end{array}\right ),\,\,\tau=-i\gamma^3\gamma^5=
\left (\begin{array}{cc}
I~,& 0\\
0~,& -I
\end{array}\right )
\label{C6}
\end{eqnarray}
with  $I$ being the unit $2\times 2$ matrix.

\section{Calculation of the $\Gamma (S)$ up to a first order in $G$}
\label{ApB}
\subsection{The case $G=0$}

In this case $\exp\left (iI_{int}\left (-i\frac{\delta}{\delta K}\right )\right )=1$, so we 
have from Eq. (\ref{036})
\begin{eqnarray}
\exp(iNW(K))&=&\exp \Big [N{\rm Tr}\ln \big (D(x,y)+K(x,y)\big )\Big ] 
\nonumber\\
\Longrightarrow W(K)&=&-i{\rm Tr}\ln \big (D(x,y)+K(x,y)\big ).
 \label{38}
\end{eqnarray}
Now, using a well-known relation, 
\begin{eqnarray}
\frac{\partial}{\partial\alpha}{\rm Tr}\ln M(\alpha)=
{\rm Tr} M^{-1}\frac{\partial M}{\partial\alpha},\label{A50}
\end{eqnarray}
where $M\equiv M(\alpha)$ is a matrix (see, e.g., Eq. (11.101) of Ref. \cite{peskin2}), 
we have from Eqs. (\ref{37}) and  (\ref{38})
$$
S^\alpha_\beta(x,y)=\frac{\delta W(K)}{\delta K_\alpha^\beta(y,x)}=
-i\int d^3sd^3t\sum_{\mu\nu}\Big [ \big (D+K\big )^{-1}\Big ]^\mu_\nu(s,t)
\frac{\delta K^\nu_\mu(t,s)}{\delta K_\alpha^\beta(y,x)}
$$\vspace{-0.5cm}
\begin{eqnarray}
=-i\int d^3sd^3t\sum_{\mu\nu}\Big [ \big (D+K\big )^{-1}\Big ]^\mu_\nu(s,t)\delta^3 (t-y)
\delta^3 (s-x) \delta_{\nu\beta}\delta_{\mu\alpha}
=-i\Big [ \big (D+K\big )^{-1}\Big ]^\alpha_\beta(x,y).
 \label{380}
\end{eqnarray}
Solving this relation with respect to $K$, we obtain
\begin{eqnarray}
K=-iS^{-1}-D. \label{038}
\end{eqnarray}
Finally, after substituting the relation (\ref{038}) into Eq. (\ref{38}) and taking into account 
the definition (\ref{0360}) of the CJT effective action $\Gamma(S)$, we have (omitting independent 
of $S$ terms) for it the following expression at $G=0$,
\begin{eqnarray}
\Gamma (S)=-i{\rm Tr}\ln \big (-iS^{-1}\big )+\int d^3xd^3y S^\alpha_\beta(x,y)D_\alpha^\beta(y,x).
 \label{0380}
\end{eqnarray}
Starting from the CJT effective action (\ref{0380}), it is possible to obtain the stationary 
equation (see Eq. (\ref{0370})) for the genuine spinor propagator $S$ of the 3-dim GN model at 
$G=0$. Taking into account the relation (\ref{A50}), it can be presented in the following form
\begin{eqnarray}
0&=&i\int d^3sd^3t\sum_{\mu\nu}\Big [S^{-1}\Big ]^\mu_\nu(s,t)\frac{\delta S^\nu_\mu(t,s)}{\delta S^\alpha_\beta(x,y)}+D_\alpha^\beta(y,x)
=i\Big [S^{-1}\Big ]^\beta_\alpha(y,x)+D_\alpha^\beta(y,x),
 \label{39}
\end{eqnarray}
where a trivial relation $\frac{\delta S^\nu_\mu(t,s)}{\delta S^\alpha_\beta(x,y)}=
\delta^3 (t-x)\delta^3 (s-y) \delta_{\nu\alpha}\delta_{\mu\beta}$ 
is taken into consideration. Hence, in the absence of interaction in the GN model (\ref{n1}), 
i.e. at $G=0$, the stable and stationary form of the propagator is the following, $S=-iD^{-1}$, where $D$ is presented in Eq. (\ref{360}).

\subsection{CJT effective action in the first order in $G$.}

In this case the functional $W(K)$ (\ref{036}) looks like (here and below we use the 
definition $\Delta\equiv D+K$)
\begin{eqnarray}
&&\exp(iNW(K))=\Big (1+iI_{int}\Big (-i\frac{\delta}{\delta K}\Big )\Big )\exp \Big (N{\rm Tr}\ln \Delta\Big )\nonumber\\
&=&\Big\{1-i\frac {G}{2N}\int d^3sd^3td^3ud^3v\delta^3 (s-t)\delta^3 (t-u)\delta^3 (u-v) \delta_\alpha^\beta\frac{\delta}{\delta K^\beta_\alpha (s,t)} \delta_\mu^\nu\frac{\delta}{\delta K^\nu_\mu (u,v)}\Big\}\exp \Big (N{\rm Tr}\ln \Delta\Big ).
\label{039}
\end{eqnarray}
In the following two relations are needed,
\begin{eqnarray}
\frac{\delta {\rm Tr} \ln\Delta}{\delta K_\mu^\nu(u,v)}=\Big ( \Delta^{-1}\Big )^\mu_\nu(v,u),
 \label{390}
\end{eqnarray}
which is a consequence of Eq. (\ref{380}) or Eq. (\ref{A50}), and
\begin{eqnarray}
\frac{\delta}{\delta K_\alpha^\beta(s,t)}\Big ( \Delta^{-1}\Big )^\mu_\nu(v,u)=
-\int d^3v'd^3u'\sum_{\mu',\nu'}\Big ( \Delta^{-1}\Big )^\mu_{\mu'}(v,v')\frac{\delta 
\Delta^{\mu'}_{\nu'}(v',u')}{\delta K_\alpha^\beta(s,t)}\Big (
\Delta^{-1}\Big )_\nu^{\nu'}(u',u). \label{0390}
\end{eqnarray}
Taking into account in the last relation that 
$\frac{\delta \Delta^{\mu'}_{\nu'}(v',u')}{\delta K_\alpha^\beta(s,t)}=\delta^3 
(v'-s)\delta^3 (u'-t)\delta^{\mu'\beta}\delta_{\nu'\alpha}$, 
we have from (\ref{0390}) that
\begin{eqnarray}
\frac{\delta}{\delta K_\alpha^\beta(s,t)}\Big ( \Delta^{-1}\Big )^\mu_\nu(v,u)=
-\Big ( \Delta^{-1}\Big )^\mu_{\beta}(v,s)\Big ( \Delta^{-1}\Big )_\nu^{\alpha}(t,u).\label{040}
\end{eqnarray}
Applying the relations (\ref{390}) and (\ref{040}) in Eq. (\ref{039}), we obtain
\begin{eqnarray}
\exp(iNW(K))=\Big\{1-i\frac{GN}2\int d^3s \Big [{\rm tr}\Delta^{-1}(s,s)\Big ]^2 +i\frac{G}2\int d^3s~ {\rm tr}\Big [\Delta^{-1}(s,s)\Delta^{-1}(s,s)\Big ]\Big\}\exp \Big (N{\rm Tr}\ln \Delta\Big ),
 \label{40}
\end{eqnarray}
where ${\rm tr}$ means the trace operation only in the spinor space. It follows from Eq. (\ref{40})
that up to a first order of $G$
\begin{eqnarray}
W(K)&=&-i{\rm Tr}\ln \Delta-\frac{G}2\int d^3s \Big [{\rm tr}\Delta^{-1}(s,s)\Big ]^2 
+\frac{G}{2N}\int d^3s~ {\rm tr}\Big [\Delta^{-1}(s,s)\Delta^{-1}(s,s)\Big ].
 \label{400}
\end{eqnarray}

To find the effective action $\Gamma(S)$ in the first order of $G$, we must use in Eq. 
(\ref{0360}), as well as in Eq. (\ref{37}), the expression (\ref{400}) for $W(K)$. In particular,
it follows from Eqs. (\ref{37}) and (\ref{400}) that
\begin{eqnarray}
S^\alpha_\beta(x,y)\equiv\frac{\delta W(K)}{\delta K_\alpha^\beta(y,x)}&=&
-i\Big ( \Delta^{-1}\Big )_\beta^\alpha(x,y)+G\int d^3s\sum_\mu\Big ( \Delta^{-1}\Big )_\mu^{\alpha}(x,s)~{\rm tr}\Delta^{-1}(s,s)\Big ( \Delta^{-1}\Big )_\beta^{\mu}(s,y)~\nonumber\\
&-&\frac{G}{N}\int d^3s\sum_{\mu\nu}\Big ( \Delta^{-1}\Big )_\mu^{\alpha}(x,s)\Big (
\Delta^{-1}\Big )_\nu^{\mu}(s,s)\Big ( \Delta^{-1}\Big )_\beta^{\nu}(s,y),
 \label{0400}
\end{eqnarray}
where the relation (\ref{040}) was applied. Now, the next problem is to express the bilocal 
sourse $K$ as a function(al) of $S$ with a help of Eq. (\ref{0400}). We will use the perturbation approarch over the coupling constant $G$, i.e., will suppose that the solution of Eq. (\ref{0400}) has the form
\begin{eqnarray}
K(S)=K_0+\delta K,
 \label{41}
\end{eqnarray}
where $\delta K\sim G$ and $K_0$ is the solution of Eq. (\ref{0400}) at $G=0$, and it is given 
in Eq. (\ref{038}), i.e., $K_0= -iS^{-1}-D$. Recall that $\Delta^{-1}$ in Eq. (\ref{0400}) is 
indeed a functional of $K$, i.e., $\Delta^{-1}\equiv \Delta^{-1}(K)$. So, let us expand this 
quantity in a Taylor series around $K_0$ up to a first order in a small perturbation 
$\delta K$ of Eq. (\ref{41}),
\begin{eqnarray}
\Big ( \Delta^{-1}(K)\Big )_\beta^\alpha(x,y)=\Big ( \Delta^{-1}(K_0)\Big )_\beta^\alpha(x,y)+\int d^3ud^3v~\delta K^\nu_\mu(u,v)\frac{\delta \Big ( \Delta^{-1}(K)\Big )_\beta^\alpha(x,y)}{\delta K^\nu_\mu(u,v)}\Big |_{K=K_0}+\cdots.
 \label{041}
\end{eqnarray}
Taking into account in Eq. (\ref{041}) the derivative rule (\ref{040}) as well as the trivial 
relation $\Big (\Delta^{-1}(K_0)\Big )_\beta^\alpha(x,y)=iS_\beta ^\alpha(x,y)$, we obtain 
\begin{eqnarray}
\Big ( \Delta^{-1}(K)\Big )_\beta^\alpha(x,y)=iS_\beta ^\alpha(x,y)+\int d^3ud^3v~S^\alpha_\nu (x,u)\delta K^\nu_\mu(u,v)S^\mu_\beta (v,y)+\cdots.
\label{410}
\end{eqnarray}
After a substitution of the relation (\ref{410}) instead of a first term in the right hand side of Eq. (\ref{0400}) and replacing all $\Delta^{-1}$ in other terms of Eq. (\ref{0400}) by $iS$, we find the following equation on the quantity
$\delta K$
\begin{eqnarray}
\int d^3ud^3v~S^\alpha_\nu (x,u)\delta K^\nu_\mu(u,v)S^\mu_\beta (v,y)&=&
-G\int d^3s\sum_\mu~S_\mu^{\alpha}(x,s)~{\rm tr}S(s,s)S_\beta^{\mu}(s,y)\nonumber\\
&+&\frac{G}{N}\int d^3s\sum_{\mu\nu}S_\mu^{\alpha}(x,s)S_\nu^{\mu}(s,s)S
_\beta^{\nu}(s,y).
 \label{0410}
\end{eqnarray}
Its solution with respect to $\delta K$ has the following form
\begin{eqnarray}
\delta K^\alpha_\beta(x,y)&=&-G\delta^\alpha_\beta\delta^3 (x-y)~{\rm tr}S(x,x)
+\frac{G}{N}S_\beta^{\alpha}(x,x)\delta^3 (x-y).
 \label{42}
\end{eqnarray}
As a result, we obtain, up to a first order in $G$, the solution $K$ of the equation (\ref{0400}),
\begin{eqnarray}
K^\alpha_\beta(x,y)=-i\Big (S^{-1}\Big )^\alpha_\beta(x,y)-D^\alpha_\beta(x,y)-
G\delta^\alpha_\beta\delta^3 (x-y)~{\rm tr}S(x,x)
+\frac{G}{N}S_\beta^{\alpha}(x,x)\delta^3 (x-y).
\label{042}
\end{eqnarray}
Substituting Eq. (\ref{042}) in a definition (\ref{0360}) of the CJT effective action, 
we have for $\Gamma (S)$ in the first order of $G$ the expression (\ref{n420}).

\end{document}